\documentstyle[twocolumn,seceq]{jpsj}
\def\geqap{\,\raise 2pt \hbox{$>\kern-11pt \lower 5pt \hbox{$\sim$}$}\,}
\def\leqap{\,\raise 2pt \hbox{$<\kern-10pt \lower 5pt \hbox{$\sim$}$}\,}
\def\runtitle{NaV$_2$O$_5$ as an Anisotropic $t$$-$$J$ Ladder at 
Quarter Filling}
\def\runauthor{Satoshi {\sc Nishimoto} and Yukinori {\sc Ohta}}
\title{NaV$_2$O$_5$ as an Anisotropic $t$$-$$J$ Ladder at Quarter Filling}
\author{Satoshi {\sc Nishimoto}$^{1}$ and Yukinori {\sc Ohta}$^{1,2}$}
\inst
{$^1$ Graduate School of Science and Technology, Chiba University, 
Inage-ku, Chiba 263-8522\\
$^2$Department of Physics, Chiba University, Inage-ku, Chiba 263-8522}
\recdate{May 21, 1998}
\abst
{Based on recent experimental evidences that the electronic charge 
degrees of freedom plays an essential role in the spin-Peierls--like 
phase transition of NaV$_2$O$_5$, we first make the mapping of 
low-energy electronic states of the $d$$-$$p$ model for NaV$_2$O$_5$ 
to the quarter-filled $t$$-$$J$ ladder with anisotropic parameter 
values between legs and rungs, and then show that this anisotropic 
$t$$-$$J$ ladder is in the Mott insulating state, of which lowest-energy 
states can be modeled by the one-dimensional Heisenberg antiferromagnet 
with the effective exchange interaction $J_{\rm eff}$ whose value is 
consistent with experimental estimates.  We furthermore examine the 
coupling between the ladders as the trellis lattice model and show 
that the nearest-neighbor Coulomb repulsion on the zigzag-chain bonds 
can lead to the instability in the charge degrees of freedom of the 
ladders.  
}
\kword
{NaV$_2$O$_5$, spin-Peierls transition, $t$$-$$J$ ladder, 
quarter filling, $d$$-$$p$ model, trellis lattice}
\begin{document}
\sloppy
\maketitle

Recently it has been reported\cite{ohama} that the phase transition 
at $T_c=34$ K in NaV$_2$O$_5$ is not a conventional spin-Peierls 
transition but rather a charge ordering into a geometrical 
configuration of charges with the spin excitation gap.  
Although there are still a lot of controversies in experimental 
data,\cite{isobe,neutron,nmr,thermal,prep1,prep2,xray,smirnov,arpes} 
this work certainly casts doubt on the present understanding of the 
basic electronic states of this system; i.e., electronic charge degrees 
of freedom cannot be eliminated in the low-energy physics of the 
system.  More specifically, we might have to start from a quarter-filled 
ladder-type model, rather than from the one-dimensional (1D) 
Heisenberg model with only the dimerizing lattice degrees of freedom, 
to discuss this phase transition.  

In this paper we consider this unconventional spin-Peierls system.  
We first analyze the low-energy states of the $d$$-$$p$ model by 
using a V$_4$O$_4$ cluster and by making one-to-one correspondence 
of the states to analytically solved states of the 4-site $t$$-$$J$ 
cluster we derive the anisotropic $t$$-$$J$ ladder Hamiltonian 
which we argue can hopefully be a starting model for the low-energy 
physics of this system.  We will show that the mapping works quite 
well and enables us to evaluate the parameter values for the 
Hamiltonian.  
A Lanczos exact-diagonalization technique is then applied to the 
finite-size clusters of this ladder model to examine its low-energy 
states.  We will thereby show that our anisotropic $t$$-$$J$ ladder 
has a finite charge gap in the realistic parameter region and the 
lowest-energy levels can be modeled by the 1D Heisenberg 
antiferromagnet with the effective exchange interaction consistent 
with experimental estimates.  
We will moreover argue that because the model still has the charge 
degrees of freedom in its low-energy sector it can be a starting 
Hamiltonian for describing the unconventional spin-Peierls system 
NaV$_2$O$_5$ if appropriate lattice degrees of freedom and/or some 
presently unknown factors are taken into account; we will suggest 
a possible scenario that a fairly small inter-ladder repulsive 
interaction can induce instability for the charge disproportionation 
in the ladders.  

High-energy physics of the system may be contained in a $d$$-$$p$ 
model cluster V$_4$O$_4$ (a pair of the rungs) which is illustrated 
in Fig.~1.  The $d$$-$$p$ Hamiltonian reads
\begin{eqnarray}
H_{d-p}=&-&\sum_{il\sigma}(\pm)^lt_{pd}
(d_{i\sigma}^\dagger p_{l\sigma}+{\rm H.c.})
+U_d\sum_i n_{i\uparrow}^d n_{i\downarrow}^d \nonumber\\
&+&\sum_i\varepsilon_i^dn_i^d+\sum_l\varepsilon_l^pn_l^p
\end{eqnarray}
where $d_{i\sigma}^\dagger$ ($p_{l\sigma}$) creates (annihilates) 
a spin-$\sigma$ electron on the $d_{xy}$-orbital at site $i$ 
(neighboring $p$-orbital at site $l$), and 
$n_{i\sigma}^d$ and $n_{l\sigma}^p$ are the electron number operator.  
We take into account 
(i) the hopping parameters $t_{pd}$ between the $d_{xy}$-orbital 
on V ion and the $p_x$-orbital ($t_{pd}^x$) and $p_y$-orbital 
($t_{pd}^y$) on the neighboring O ions where the phase factor 
$(\pm)^l$ has to be taken appropriately, 
(ii) energy-level difference $\Delta_{pd}=\varepsilon_i^d-\varepsilon_l^p$ 
between the $d_{xy}$-orbital and the $p_x$-orbital ($\Delta_{pd}^x$) 
and $p_y$-orbital ($\Delta_{pd}^y$), and 
(iii) Coulomb repulsion $U_d$ on V ions.  
The anisotropy of the parameter values will turn out to be 
important.  We note that this cluster is the smallest one in that 
it takes into account the anisotropic legs and rungs simultaneously; 
i.e., it contains the effective repulsive interaction on the rungs 
which can lead to the charge gap, the effective exchange interaction 
between the rungs which can lead to the 1D Heisenberg antiferromagnet, 
and the exchange interaction between the neighboring V ions on the 
leg of the ladder which can lead to the spin gap when the system is 
dimerized at $T<T_c$.  
The coupling between ladders is neglected for the moment 
(see below for this effect).  
Values of the $d$$-$$p$ model parameters are taken from 
Ref.\cite{horsch}: 
$t_{pd}^x$ and $t_{pd}^y$ take the value of $1.22$ and $1.03$ eV, 
respectively, but are assumed to depend on the bond length $d$ 
as $t_{pd}\propto d^{-7/2}$,\cite{harrison}  
$\Delta_{pd}^x=6.5$ eV, $\Delta_{pd}^y=4$ eV, and $U_d=4$ eV.  
There are 10 electrons in the cluster.  
\begin{figure}
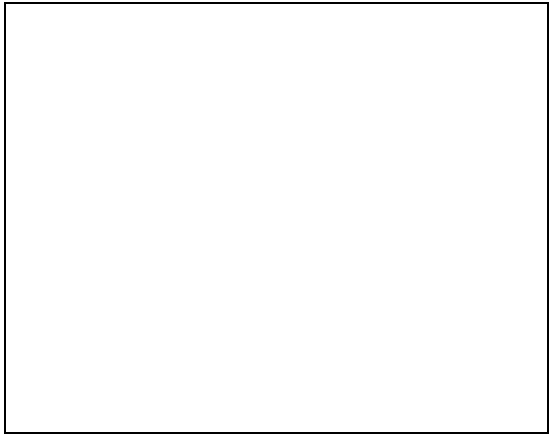

\figureheight{5.5cm}
\caption{Orbital structure of (a) the V$_2$O$_3$ ladder 
in NaV$_2$O$_5$ and (b) a corresponding $t$$-$$J$ ladder.  
The small clusters used for the mapping are indicated by 
dotted lines.}
\label{fig:1}
\end{figure}

Energies of several low-lying states of the V$_4$O$_4$ cluster 
are evaluated by a Lanczos diagonalization method and are compared 
with the analytically solved low-energy states of the 4-site 
$t$$-$$J$ cluster with two electrons.  
The $t$$-$$J$ Hamiltonian reads 
\begin{eqnarray}
H_{t-J}=&-&\sum_{\langle ij\rangle\sigma}t_{ij}
\big(\hat{c}^\dagger_{i\sigma}\hat{c}_{j\sigma}+{\rm H.c.}\big)
\nonumber \\
&+&\sum_{\langle ij\rangle}J_{ij}\Big({\bf S}_i\cdot{\bf S}_j
-\frac{1}{4}n_in_j\Big)
\end{eqnarray}
where $\langle ij\rangle$ represents the nearest-neighbor 
bonds along the legs and rungs with the hopping and exchange parameters 
$t_{ij}$ and $J_{ij}$ taking $t_\perp$ and $J_\perp$ for the rungs 
and $t_\parallel$ and $J_\parallel$ for the legs.  
$\hat{c}^\dagger_{i\sigma}=c^\dagger_{i\sigma}(1-n_{i-\sigma})$ 
is the constrained electron-creation operator at site $i$ and 
spin $\sigma$ $(=\uparrow,\downarrow)$, ${\bf S}_i$ is the 
spin-$\frac{1}{2}$ operator, and $n_i=n_{i\uparrow}+n_{i\downarrow}$ 
is the electron-number operator.  
We confirm that the several low-lying states of both $d$$-$$p$ and 
$t$$-$$J$ clusters have the same quantum numbers and characters and 
thus we find that the mapping is straightforward.  Here we determine 
the values of the 4 ladder-parameters by fitting the energies of the 
lowest 5 states for the $d$$-$$p$ and $t$$-$$J$ clusters (the extra 
one is for the origin of energy).  The obtained values of the ladder 
parameters for the realistic bond lengths are: 
$t_\perp=0.298$, 
$t_\parallel=0.140$, 
$J_\perp=0.049$, and 
$J_\parallel=0.029$ in units of eV 
(which we will use in the following discussions unless otherwise 
indicated).  
In Fig.~2, we show the bond-length dependence of these parameters.  
We note that with increasing $d_\parallel$, the value 
of $J_\parallel$ decreases as expected, but the value $J_\perp$ 
also decreases considerably.  Also noted is that with increasing 
$d_\perp$ the value of $J_\parallel$ increases.  These are from 
the higher-order contributions to the exchange interactions.  
The hopping parameters $t_\perp$ and $t_\parallel$ on the other 
hand vary more or less as expected.  
\begin{figure}
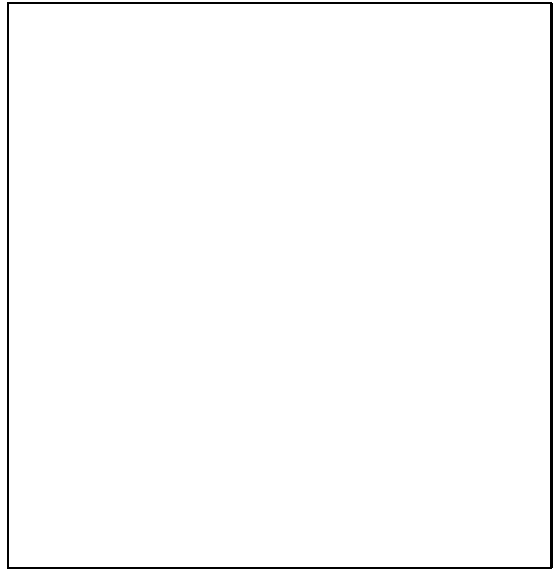

\figureheight{7.3cm}
\caption{Bond-length dependence of the estimated parameter values 
for the $t$$-$$J$ ladder.  
In (a) and (b) the V-O bond length on the rungs $d_\perp$ is kept 
constant at $d_\perp=1.825$ \AA, while in (c) and (d) the V-O 
bond length on the legs $d_\parallel$ is kept constant at 
$d_\parallel=1.916$ \AA.  
}
\label{fig:2}
\end{figure}

The single-particle spectra from the quarter-filled ground states 
of the V$_4$O$_4$ cluster and corresponding $t$$-$$J$ cluster are 
shown in Fig.~3.  We find a very good agreement, demonstrating that 
the mapping is satisfactory also for the one-electron added 
and one-electron removed sectors of the cluster Hilbert space.  
\begin{figure}
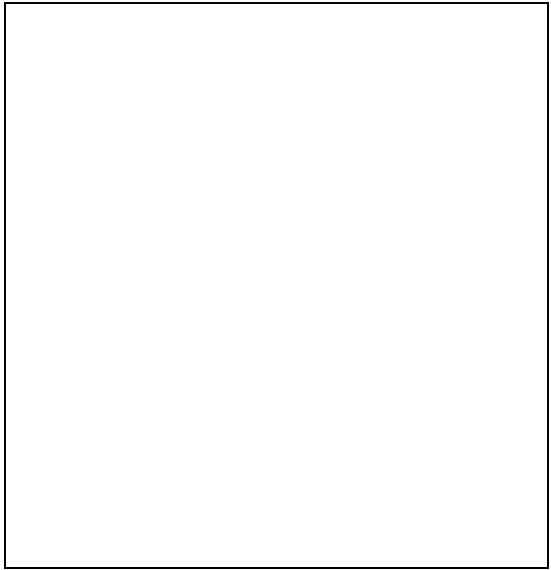

\figureheight{7.3cm}
\caption{Single-particle spectra for the $d$$-$$p$ cluster 
compared with those of the $t$$-$$J$ cluster at quarter 
filling.  The spectra near the chemical potential (set at 
$\omega=0$) are enlarged in the lower panel.  
For the $d$$-$$p$ model, the total spectra summing up the 
contributions from the V $d_{xy}$-orbital and O $p_x$- and 
$p_y$-orbitals are shown although around the chemical 
potential the spectral weight is almost completely from 
the V $d_{xy}$-orbital.  
}
\label{fig:3}
\end{figure}

Now let us examine the electronic states of the obtained 
anisotropic $t$$-$$J$ ladder at quarter filling.  
Here we examine the charge gap $\Delta_c$ of the model, which 
is not yet well known although there were some 
discussions.\cite{riera}  The charge gap is defined by 
\begin{eqnarray}
\Delta_c=\frac{1}{2}\Big[\big[E_0(N_\uparrow+1,N_\downarrow)
&-&E_0(N_\uparrow,N_\downarrow)\big]   \nonumber \\
-\big[E_0(N_\uparrow,N_\downarrow)
&-&E_0(N_\uparrow-1,N_\downarrow)\big]\Big] 
\end{eqnarray}
where $E_0(N_\uparrow,N_\downarrow)$ is the ground-state energy 
of the system with $N_\uparrow$ up-spin and $N_\downarrow$ down-spin 
electrons.  This is evaluated for 8, 12, and 16-site clusters 
with either periodic or antiperiodic boundary condition to 
minimize the shell effect on the values of $\Delta_c$.  
The results are plotted in Fig.~4 as a function of $1/L$ 
($L$ is the length of the ladder) to see the value at the 
infinite system size.  
We find that at least for $t_\perp/t_\parallel\geqap 1.3-1.5$ 
the gap opens up although the situation is still unclear for 
$t_\perp\simeq t_\parallel$ in the present calculations of 
small finite-size systems.  We also calculate the equal-time charge 
correlation function $\langle n_in_j\rangle$ and find that 
the probability of finding two electrons on the same rung is 
very small for $t_\perp/t_\parallel\geqap 1.3$, indicating 
that one electron is localized on each rung.  It seems thus quite 
reasonable to say that the anisotropic ladder at quarter 
filling is in the insulating state for the realistic parameter 
values.  
\begin{figure}
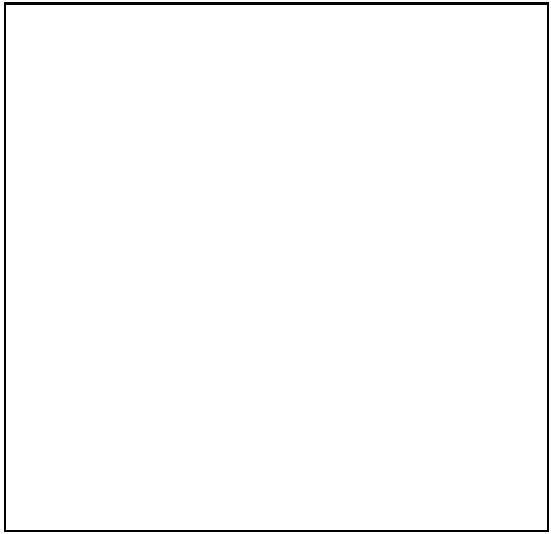

\figureheight{6.8cm}
\caption{Charge gap $\Delta_c/t_\parallel$ of the finite-size 
clusters for the quarter-filled $t$$-$$J$ ladder as a function 
of the inverse of the cluster size $1/L$.  Anisotropy in the 
hopping parameters $t_\perp/t_\parallel$ is varied with 
keeping $J_\perp/J_\parallel=2$ and $J_\perp/t_\perp=0.2$ 
constant.  
}
\label{fig:4}
\end{figure}

The mechanism of this charge localization is common in a class of 
quarter-filled organic systems where the Mott insulating phase is 
realized due to dimerization: with increasing dimerization 
strength a dimer turns into an effective single site of a 
half-filled antibonding state with Hubbard-like repulsive 
interaction.\cite{kino,seo,nishimoto1,nishimoto2}  
This mechanism operates also in the anisotropic ladder at quarter 
filling if we can regard the rung as a dimer, and actually in some 
parameter region, we do have this insulating phase as we have seen 
above.  This Mott insulating phase can be modeled as a 1D Heisenberg 
antiferromagnet of which the effective exchange interaction 
is estimated to be $J_{\rm eff}=85$ meV from the singlet-triplet 
splitting in the 4-site $t$$-$$J$ cluster.  This value is consistent 
with the value obtained in Ref.\cite{horsch} and should also be 
consistent with values determined from the observed temperature 
dependence of the uniform susceptibility $\chi (T)$ if a small 
inter-chain exchange coupling and some other effects are taken 
into account.\cite{horsch}  
Note that the value of $J_\parallel$ is much smaller than the 
value of $J_{\rm eff}$ (see Fig.~3); the picture that 
the electrons order on one of the legs of the ladder already 
above $T_c$ should predict a very different peak position in 
$\chi(T)$.  Also noted is that the bond-length dependence of 
$J_{\rm eff}$ is very strong and anisotropic (see Fig.~3), which 
may be confirmed by some high-pressure experiment.  
The resultant picture for the high-temperature phase of 
NaV$_2$O$_5$ is thus equivalent to that of Horsch and Mack\cite{horsch} 
although the procedure of derivation of the 1D Heisenberg 
model is quite different.  

In the real materials, we have to take into account the coupling 
between the ladders, i.e., the system may be the one modeled as a 
two-dimensional (2D) trellis lattice (ladders connected with zigzag 
chain bonds as in Fig.~1 (b)).  
We examine the electronic states, i.e., the equal-time charge 
correlation function and charge gap for the 16-site clusters of 
this lattice (or coupled two $4\times 2$ ladders), and find that 
the basic electronic state that one electron is localized on 
each rung is maintained against the inter-ladder hopping integral 
of up to $t_{xy}\simeq 1.2 t_\parallel$.  The reported values of 
$t_{xy}$ are fairly small, so that the electronic state of the 
real materials may be in this localization regime.\cite{xray,horsch}  
We note that the situation is similar to the case in the 2D 
dimerized lattice system of $\kappa$-(BEDT-TTF)$_2$X where the Mott 
insulating state is realized over a wide range of parameter 
values.\cite{nishimoto2}  

We now want to consider the low-temperature phase of NaV$_2$O$_5$, 
i.e., the mechanism of charge ordering reported in Ref.\cite{ohama}.  
This is in principle possible because our starting Hamiltonian, 
the $t$$-$$J$ ladder at quarter filling, still has the charge 
degrees of freedom in the low-energy regions, unlike in the coupled 
Heisenberg chain in Ref.\cite{horsch}.  
We however immediately notice that it seems unlikely that the 
$t$$-$$J$ ladder with the small $J/t$ values has inherent charge 
instability in itself even though it is anisotropic; the only 
possibility seems to be the phase separation realized for very 
large $J_\parallel$ values where the electrons are localized on 
one of the legs to form a Heisenberg chain.  

The simplest interaction that can lead to the charge instability 
is the intersite Coulomb repulsion.  In (TMTTF)$_2$X, e.g., the 
nearest-neighbor Coulomb repulsion is discussed to be the origin 
of the observed charge disproportionation.\cite{seo}  
Here we introduce the repulsive interaction $V$ between ladders, 
i.e., on the zigzag chain bonds running along the crystallographic 
$b$-axis, which have the shortest bond-length among V-V bonds.  
We use the two $4$$\times$$2$-ladder clusters for modeling 
the trellis lattice and calculate the charge correlation 
function $\langle n_in_j\rangle$.  As shown in Fig.~5, the tendency 
to the charge disproportionation where the electrons are localized 
on one side of the legs is actually found for not too large values 
of $V$ (only 10--20\% of $U_d$).  Once the 1D Heisenberg chain on 
one of the legs of the ladder is thus realized (by some phase 
transition in the thermodynamic limit) the system may be subject 
to the lattice dimerization which leads to the opening of the spin 
excitation gap.  
\begin{figure}
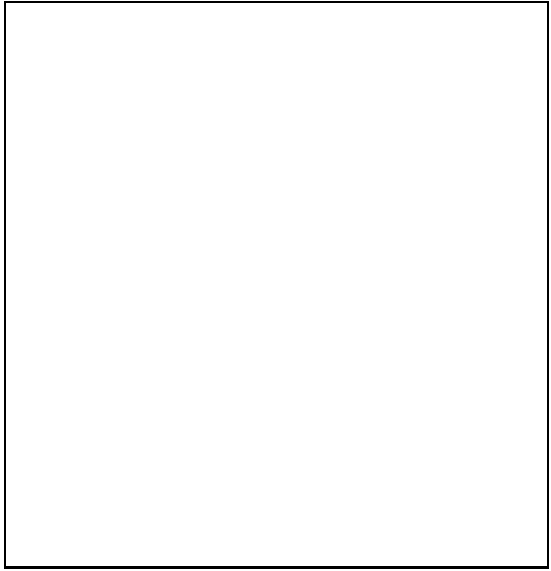

\figureheight{7.3cm}
\caption{Charge correlation function $\langle n_in_j\rangle$ 
for the $t$$-$$J$ trellis lattice as a function of the 
density repulsion $V$ between the ladders.  
The site indeces defined in Fig.~1 (b) are used with $i=1$.  
The parameter values for the zigzag chain bond are assumed 
as $t_{xy}=t_\parallel$ and $J_{xy}=J_\parallel$.  
}
\label{fig:5}
\end{figure}

An optimistic speculation for the phase transition of 
NaV$_2$O$_5$ is a realization of this charge ordering.  
In this picture, the critical temperature of the charge ordering 
needs not be the same as the spin-Peierls transition temperature 
of the spin chain of one of the legs of the ladder; if the former 
temperature is higher than the latter, we should observe 
two transitions independently.  However the opposite seems to be 
the case in real materials: when the charge ordering occurs, 
the spin chain of one of the legs of the ladder is already 
below its spin-Peierls transition temperature, so that the chain 
dimerizes simultaneously at $T_c$ into the dimerization geometry 
potentially prepared at a higher temperature.  A significant deviation 
from the BCS-type temperature dependence of the gap function\cite{neutron} 
might be attributed to this.\cite{ohama2}  
Also interesting is a report from mean-field calculations 
of the 1D extended Hubbard model for (TMTTF)$_2$X where it is 
shown that the first-order phase transition of the charge ordering 
crosses over to a second-order--like behavior when the ratio 
of the values of two hopping parameters deviates from 1,\cite{seo} 
which is suggestive to the observed second-order--like phase 
transition of NaV$_2$O$_5$.\cite{isobe}  
However, to decide if the mechanism suggested above really applies, 
we of course have to know much about the electronic states of this 
system from further experimental studies.  

In summary, we have studied the electronic states of NaV$_2$O$_5$ 
by making the mapping of low-energy electronic states of the 
$d$$-$$p$ model for this compound to the quarter-filled $t$$-$$J$ 
ladder with anisotropic parameter values between legs and rungs.  
We have shown that this anisotropic $t$$-$$J$ ladder is in 
the Mott insulating state, of which lowest-energy states can be 
modeled by the one-dimensional Heisenberg chain with the effective 
exchange interaction $J_{\rm eff}$ whose value is consistent with 
experimental estimates.  We have also studied the coupling between 
the ladders by using the trellis lattice model and have shown that 
the intersite Coulomb repulsion on the zigzag-chain bonds can leads 
to the instability in the charge degrees of freedom of the ladders.  
A possible mechanism of the spin-Peierls--like phase transition of 
NaV$_2$O$_5$ has thereby been suggested.  

We would like to thank T. Ohama and H. Sawa for enlightening 
discussions on experimental aspects of NaV$_2$O$_5$.  
Financial supports of S.~N. by Sasakawa Scientific Research 
Grant from the Japan Science Society and of Y.~O. by Iketani Science 
and Technology Foundation are gratefully acknowledged.  
Computations were carried out in Computer Centers of the Institute 
for Solid State Physics, University of Tokyo and the Institute for 
Molecular Science, Okazaki National Research Organization.

\end{document}